\documentclass[final,3p,times,twocolumn,nopreprintline]{elsarticle}
\usepackage{hyperref}

\usepackage[utf8]{inputenc}
\usepackage[T1]{fontenc}

\usepackage{amsmath,amssymb}
\usepackage{siunitx}
\usepackage{physics}

\newcommand{\euler}{\text{e}}

\usepackage{graphicx}
\usepackage{color}

\usepackage{natbib}
\setcitestyle{square,numbers}

\usepackage{booktabs}

\usepackage{listings}
\begin{document}

\begin{frontmatter}

\title{Multisine electrochemical impedance spectroscopy for Li-ion battery characterisation}

\author[inst1,inst3]{Noël Hallemans}
\author[inst2]{Heiko Seel-Mayer}
\author[inst2]{Peter Keil}
\author[inst1,inst3]{Stephen R. Duncan}
\author[inst1,inst3]{David Howey\corref{cor1}}
\cortext[cor1]{Corresponding author}
\ead{david.howey@eng.ox.ac.uk}

\affiliation[inst1]{
            organization={Department of Engineering Science, University of Oxford},
            city={Oxford},
            postcode={OX1 3PJ},
            country={UK}
            }
\affiliation[inst2]{
            organization={Battery Dynamics},
            city={Garching},
            postcode={85748},
            country={Germany}
            }
\affiliation[inst3]{
            organization={The Faraday Institution},
            addressline={Harwell Campus},
            city={Didcot},
            postcode={OX11 0RA},
            country={UK}
            }

\begin{abstract}
Electrochemical impedance spectroscopy (EIS) is a valuable tool for non-invasive battery characterisation, providing a compact representation of physical processes over a wide range of time scales. Commonly, sinusoids at different frequencies are injected sequentially (single-sines). Alternatively, a multisine excitation (a sum of sines) is advantageous for reducing experiment time and allowing impedance to be measured during operation. In this work, we demonstrate high-fidelity multisine EIS measurements on Li-ion cells over a wide frequency range (\SI{20}{mHz} to \SI{1}{kHz}), taken with a commercial potentiostat modified to give access to the underlying current and voltage time-series data, and compare these to single-sine EIS. We show how this allows the conditions of linearity and stationarity to be verified directly, which is not possible with impedance data alone. We then measure broadband impedance during charging and discharging, temperature change, and relaxation. Because impedance is a linearisation of the response about an operating point, each of these conditions probes a state that classical EIS excludes by definition: kinetics linearised about a non-zero current, thermally activated reaction rates, and relaxing concentration gradients, respectively. Operando multisine EIS is therefore not simply a faster measurement approach, but one that resolves behaviour that steady-state EIS cannot.
\end{abstract}
\end{frontmatter}

\section{Introduction}

Electrochemical impedance spectroscopy (EIS) \cite{lasia2002electrochemical,EISbook,wang2021electrochemical,lazanas2023electrochemical} is a non-invasive technique for characterising electrochemical system dynamics. The most common approach is to apply a small-amplitude current (or voltage) sinusoid to a system, then measure the response and calculate the impedance, repeating this over several frequencies \cite{ciucci2019modeling}. Although this technique has become increasingly popular in recent years, computer-controlled EIS has been around since at least the 1980s---for example, Boukamp built a microcomputer-based wideband measurement system in 1984 \cite{boukamp1984microcomputer}. Early applications of EIS were mainly for examining corrosion, anodising, and coatings \cite{mansfeld1988electrochemical,van1990investigation}, but it was not long before it was also used for characterising battery electrodes \cite{levi1997simultaneous}. Impedance is particularly attractive for battery characterisation since it provides a compact data representation for identifying physical processes whose characteristic time scales may be orders of magnitude apart \cite{ciucci2019modeling,gabervsvcek2021understanding}. Today, EIS is routinely applied in industry and academia for estimating battery health \cite{hu2023application}, parametrising models \cite{bizeray2018identifiability,hallemans2025physics,hileman2024estimating}, monitoring temperature \cite{li2022temperature, richardson2014battery}, investigating the performance of battery materials \cite{hein2020influence,drummond2022modelling,menkin2024insights,matthews2025impact}, and much more.

Classical impedance measurements must satisfy the conditions of \textit{linearity}, \textit{stationarity}, and \textit{causality} \cite{urquidi1990applications,hirschorn2008selection,hallemans2023electrochemical,goh2024comparison}, and hence are limited to providing only linearised small-signal snapshots of the behaviour of systems that in reality are nonlinear and nonstationary. In the case of batteries, for example, this means probing the response at fixed state-of-charge (SOC) and temperature, after relaxation. Before analysing impedance data, the conditions of linearity, stationarity, and causality should be verified, for instance via Kramers-Kronig relations \cite{urquidi1990applications,esteban1991application,boukamp2025guidance}, Lissajous plots \cite{zabara2024utility}, or total harmonic distortion \cite{giner2015total}.

\begin{figure}[t]
    \centering
    \includegraphics[width=\columnwidth]{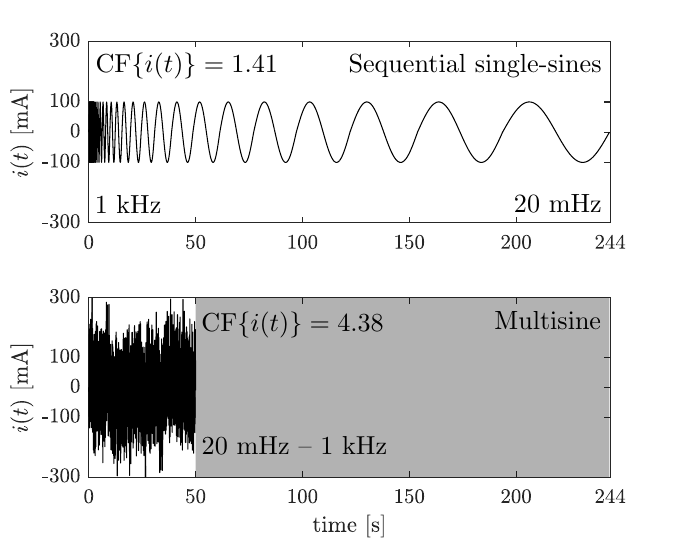}
    \caption{Single-sine and multisine current excitation signals for battery tests, both with amplitude \SI{70.5}{mA}~rms. The multisine is shorter (\SI{50}{s} vs.\ \SI{244}{s} for single-sine) but has higher peak currents.}
    \label{Fig:singleSineMultisine}
\end{figure}

While EIS is generally performed with single-sine excitation (with sinusoids injected \textit{sequentially}), an alternative approach is to apply several frequencies \textit{simultaneously} by summing sinusoids together, which is referred to as a multisine signal \cite{hallemans2023electrochemical,van2009advantages,widanage2016design,ulgut2022methods,kallel2023design,fan2025fast}. Single-sine excitation can result in long measurement times, depending on the frequency range and number of frequencies per decade, whereas multisine excitation has the advantage of only being limited by the lowest excitation frequency, Fig.~\ref{Fig:singleSineMultisine}. A disadvantage is that constructive interference can yield larger perturbation amplitudes that may cause nonlinear responses.

Multisine excitation allows measurement of broadband impedance data under operationally realistic conditions  \cite{koster2017dynamic,zappen2018application,zappen2020operando,zhu2022operando,hallemans2022operando}, which is useful for exploring application-specific behaviour. For batteries in the field, this offers the opportunity to monitor impedance \textit{in operando} \cite{zheng2024real,sihvo2025real,hackmann2025operando,kirst2025plating}, for example during electric vehicle charging. In addition to automotive applications, operando multisine EIS also enables novel approaches for materials characterisation in laboratories, for instance studying solid-electrolyte interphase formation in anode-free Li-metal cells \cite{dabiri2025operando}.

The problem with using single-sines for operando measurements is that, by injecting sinusoids sequentially instead of simultaneously, every frequency is applied at a different operating condition. This is only valid at high frequencies (i.e., short excitation periods) on slowly-varying systems \cite{drvarivc2025operando}, and is therefore limited in applicability. There is also a more fundamental motivation for operando measurement. Because classical EIS requires equilibrium, it necessarily usually linearises the response about zero current, where the kinetics are symmetric by construction. The charge-transfer resistance obtained this way therefore reflects the exchange current density alone, and carries no information about the asymmetry of the kinetics between charging and discharging. Measuring impedance at a non-zero operating current removes this restriction, and gives access to parameters that are not identifiable from steady-state data.

Although multisine EIS is a powerful tool with several possible applications, commercially available potentiostats are typically provided with only a limited multisine frequency range, and the directly measured current and voltage data are usually not available to users. There are practical reasons for this, such as memory and bandwidth constraints, but these limitations mean that the full power of multisine cannot be exploited. In this paper, we use a modified commercial potentiostat (Battery Dynamics Series S) to measure multisine current and voltage data, and discuss how to use this data to compute impedance. We show that single-sine and multisine EIS give the same results in steady-state, and discuss in detail how linearity and stationarity can be verified. We show that applying single-sines during operation can lead to erroneous impedance measurements, and that multisine excitation resolves this issue. We then measure multisine EIS on Li-ion batteries over several relevant operating conditions (during charging and discharging, relaxation, and temperature changes), providing a dataset for battery characterisation and demonstrating that impedance measured during operation differs from its steady-state counterpart in ways that carry physical information rather than representing measurement error. Lastly, we give guidance on choosing between single-sine and multisine excitation depending on the application.

For both single-sine and multisine excitation, the measured current and voltage data contains additional information about linearity, stationarity, causality, and uncertainty that can be used for further analysis (e.g.\ for nonlinear EIS \cite{kirk2023nonlinear,ji2023second,ji2024second,ulrich2025early}) and it would be beneficial to researchers if potentiostat manufacturers could make this data available.

\section{Multisine design}

A multisine signal is a sum of sinusoids at different frequencies \cite{schoukens1988survey},
\begin{align}
u(t)= \sum_{m=1}^M A_m\sin(\omega_m t+\varphi_m),
\label{eq:multisine}
\end{align}
with $M$ being the number of frequencies, $A_m$ the amplitudes, $\omega_m=2\pi f_m$ the angular frequencies, $f_m$ the frequencies, and $\varphi_m$ the phases of the sinusoids. For $M=1$ and $\varphi_m=0$ we obtain a single-sine excitation.

\paragraph{Frequencies} An important constraint for multisine EIS is that all excited frequencies must be integer multiples of the fundamental (typically the lowest frequency $f_1$) for the signal to be periodic, that is,
\begin{align}
    f_m=h_m f_1 \quad \text{for} \quad h_m \in \mathbb{H}_\mathrm{exc}\subset\mathbb{N}_0,
\end{align}
with $h_m$ being the integer harmonics of the multisine and $\mathbb{H}_\mathrm{exc}$ the set of excited harmonics with cardinality $M$. The largest excited frequency is $f_\mathrm{max}=h_M f_1$ and the period of the multisine is $T_\text{p}=1/f_1$.

Accordingly, the measurement time depends only on the lowest frequency, and is shorter than for a single-sine experiment covering the same frequency range, as shown in  Fig.~\ref{Fig:singleSineMultisine}. Moreover, for a multisine we only need to wait once  for transient effects to fade out, while for a single-sine experiment we need to wait for transients to fade out at every applied frequency \cite{van2009advantages}. This results in measurement times
\begin{align}
T_\text{multisine}&=\frac{1}{f_1} + T_\text{transients}(f_1),\\
T_\text{single-sine}&=\sum_{m=1}^M \left( \frac{1}{f_m} + T_\text{transients}(f_m) \right),
\end{align}
where $T_\text{transients}(f_m)$ is the time required for transients of frequency $f_m$ to fade out. Note that techniques exist to remove the effect of transients in measurements \cite{hallemans2021best}.

The choice of excited frequency range $[f_1,f_\mathrm{max}]$ should be based on the physical processes one wishes to study. In the battery context, low frequencies provide information about diffusion while higher frequencies relate to charge transfer kinetics, double layer capacitance, and internal resistance \cite{hallemans2025physics}. Here we measure impedance in the band [\SI{20}{mHz}, \SI{1}{kHz}], and space frequencies logarithmically with ten per decade, however, compromising resolution at low frequencies due to the requirement for harmonics to be integer multiples of one another (i.e., there are fewer than 10 per decade at lower frequencies). This logarithmic spacing aims to weight the different physical processes in the impedance data evenly.

Nonlinearities in electrochemical devices result in frequencies at integer multiples of the fundamental \cite{hallemans2023electrochemical}. Hence, it is useful to design the set of excited frequencies $\mathbb{H}_\mathrm{exc}$ so as to leave ``traps'' (unexcited harmonics) for nonlinear distortions to be detected. In this work we only excite \textit{odd} harmonics and leave one in seven consecutive odd harmonics unexcited, allowing even and odd nonlinear distortions to be distinguished \cite{hallemans2023electrochemical,van2009advantages,hallemans2022operando}.

\begin{figure}[t]
    \centering
    \includegraphics[width=\columnwidth]{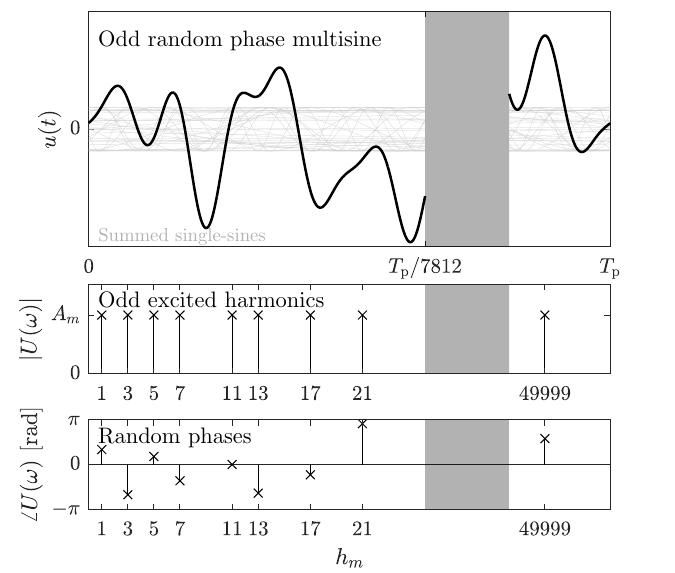}
    \caption{Odd random-phase multisine signal. The grey signals are the $M=40$ single-sines that are summed together to obtain the multisine (black). Only the beginning and end of the time- and frequency-domain signals are plotted.}
    \label{Fig:designMultisine}
\end{figure}

\paragraph{Phases} The phases $\varphi_m$ of the sinusoids should be chosen to distribute the power of the multisine uniformly in time, as  quantified by the crest factor
\begin{align}
    \mathrm{CF}\{u(t)\}=\frac{\max u(t)}{\sqrt{\frac{1}{T_\text{p}}\int_0^{T_\text{p}}u(t)^2 \mathrm{d}t}},
    \label{eq:crestfactor}
\end{align}
indicating the peak-to-rms (root-mean-square) ratio of the signal. For a fixed maximum value, a signal with lower crest factor injects more power into the system \cite[Section 5.2]{pintelon2012system}. A single-sine excitation has a crest factor of $\sqrt{2}$, while a multisine with the same rms value exhibits a higher crest factor (see Fig.~\ref{Fig:singleSineMultisine}). Summing sinusoids with the same phase typically leads to constructive interference and an increase in crest factor. To avoid this, multisine signal phases $\varphi_m$ may be optimised to minimise the crest factor \cite{guillaume1991crest,kallel2022crest}. However, random phases drawn from a uniform distribution between $[0,2\pi)$ also provide a near-optimal solution to minimising the crest factor \cite{van2009advantages,kallel2023design}, and we use them in this work. We refer to a multisine signal with randomised phases and odd harmonic excitation as an \textit{odd random-phase multisine} \cite{van2009advantages}, as illustrated in Fig.~\ref{Fig:designMultisine}.

\paragraph{Amplitudes} The excitation amplitudes $A_m$ in linear EIS are a trade-off between achieving a good signal-to-noise ratio (SNR) vs.\ the device having a linear response. The larger the amplitudes, the better the SNR, although this also strongly depends on the accuracy and design of the measurement equipment. Very high amplitudes, however, cause nonlinear behaviour \cite{hallemans2023electrochemical}. A battery is typically closest to being nonlinear at low frequencies due to the nonlinear open circuit voltage (OCV) function versus charge state.

\section{Measurement setup}
\begin{figure*}
    \centering
    \includegraphics[width=\textwidth]{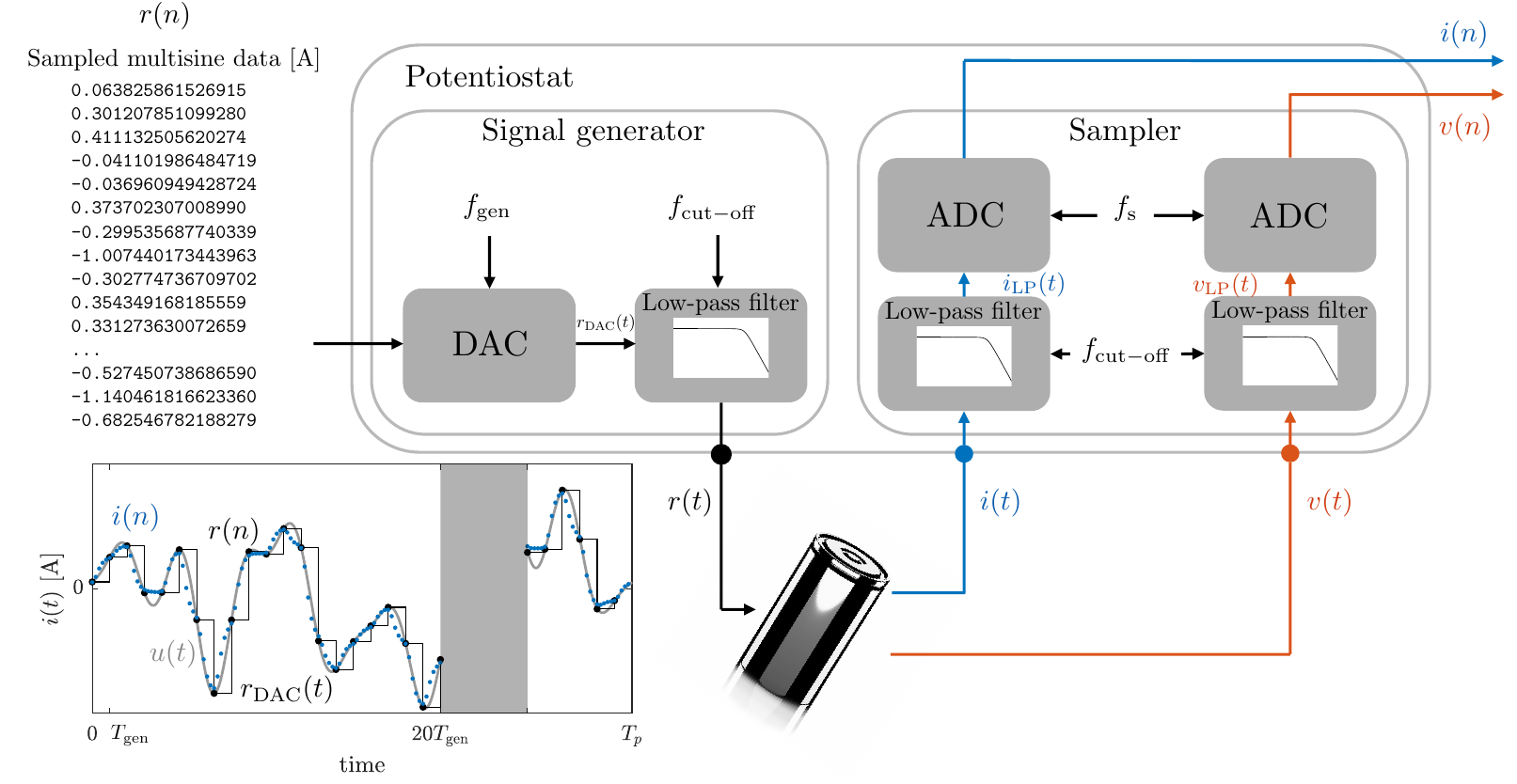}
    \caption{Potentiostat generation and measurement stages for multisine impedance. A reference current is applied to the battery and the current and voltage are measured. Continuous multisine signal $u(t)$ (grey), multisine data $r(n)$ with time step $T_\text{gen}$ (black dots), continuous signal $r_\mathrm{DAC}(t)$ after DAC (black), and sampled data $i(n)$ (blue); in practice the sampled current would be slightly delayed versus $u(t)$, but we have drawn it aligned here for clarity.}
    \label{Fig:potentiostat}
\end{figure*}

We now describe how to apply a multisine signal to a battery and measure current and voltage data with a potentiostat. We used a commercial instrument (Battery Dynamics High Resolution Tester Series S) that delivers currents of \SI{-1}{A} to \SI{1}{A}. The different components of the potentiostat are shown in Fig.~\ref{Fig:potentiostat} and described in the following paragraphs.
In this work we measured impedance under galvanostatic conditions (i.e., current-controlled), which is recommended for safety in low-impedance devices such as batteries because voltage-control can cause very high currents to flow.

\subsection{Signal generation}
First, discrete multisine data are sampled from the continuous signal $u(t)$ \eqref{eq:multisine},
\begin{align}
        r(n)=u(nT_\mathrm{gen}) \qquad n=0,1,\hdots,N_\mathrm{gen}-1,
\end{align}
where $T_\mathrm{gen}$ is the time step size and $N_\mathrm{gen}=T_\mathrm{p}/T_\mathrm{gen}$ is the number of data points (the period $T_\mathrm{p}$ should be an integer multiple of $T_\mathrm{gen}$). The generator frequency $f_\mathrm{gen}=1/T_\mathrm{gen}$ should be larger than the largest frequency of the multisine $f_\mathrm{max}$. In this work we excited frequencies up to $f_\mathrm{max}=1$~kHz and choose $f_\mathrm{gen}=3.125$~kHz (see Table~\ref{Table:measurementParameters}). The continuous multisine $u(t)$ and discrete reference data $r(n)$ are illustrated in Fig.~\ref{Fig:potentiostat}. Note that $N_\mathrm{gen}$ data points must be stored in the potentiostat to generate the excitation. The wider the frequency band measured, the larger $N_\mathrm{gen}$, and hence the larger the required memory. This is often a bottleneck for broadband multisine impedance measurements.

The potentiostat generates a continuous signal from the discrete multisine data $r(n)$ through a digital-to-analog converter (DAC). A zero-order-hold output is used (illustrated by the ``staircases'' in Fig.~\ref{Fig:potentiostat}), followed by a low-pass reconstruction filter with cut-off frequency \SI{1.6}{kHz} to limit high-frequency artefacts, resulting in a continuous signal $r(t)$ applied as the battery current through a drive circuit.

\begin{table}[t]
\centering
\begin{tabular}{lcl} \hline
Name & Parameter & Value\\
\hline
Period length& $T_\mathrm{p}$ & \SI{50}{s} \\
Fundamental frequency & $f_1$ & \SI{20}{mHz}\\
Largest frequency & $f_\mathrm{max}$ & \SI{1}{kHz}\\
Number of frequencies & $M$ & 40\\
\hline
Generator frequency & $f_\mathrm{gen}$ & \SI{3.125}{kHz}\\
Generator period & $T_\mathrm{gen}$ & \SI{320}{\micro s}\\
Sequence length & $N_\mathrm{gen}$ & 156250 \\
\hline
Sampling frequency & $f_\mathrm{s}$ & \SI{15.625}{kHz}\\
Sampling period & $T_\mathrm{s}$ & \SI{64}{\micro s}\\
Samples per period & $N_\mathrm{p}$ & 781250\\
\hline
\end{tabular}
\caption{Measurement parameters for the multisine experiments.}
\label{Table:measurementParameters}
\end{table}

\subsection{Measurement}
To calculate the impedance, the current $i(t)$ and voltage $v(t)$ are measured and then uniformly sampled in time using an analog-to-digital converter (ADC) with sampling frequency $f_\mathrm{s}$. Anti-aliasing low-pass filters should be applied to the measurements before sampling in order to limit their bandwidth.
Here, anti-aliasing filters with cut-off frequency $f_\mathrm{cut-off}=65$~kHz were used.

We chose the sampling frequency to be an integer multiple of the generator frequency ($f_\mathrm{s}=5f_\mathrm{gen}$), which leads to a total oversampling factor ($f_\mathrm{s}/f_\mathrm{max}$) of  approximately 16, reducing the effect of aliasing further. The measured current and voltage data are
\begin{align}
    & i(n)=i_\mathrm{LP}(n T_\mathrm{s}) & & n=0,1,\hdots,N-1 \nonumber\\
    & v(n)=v_\mathrm{LP}(n T_\mathrm{s}) & & n=0,1,\hdots,N-1,
    \label{eq:currentVoltageData}
\end{align}
with $i_\mathrm{LP}(t)$ and $v_\mathrm{LP}(t)$ the low-pass filtered signals, $T_\mathrm{s}=1/f_\mathrm{s}$ the sampling time,  $N=P T_\mathrm{p}f_\mathrm{s}$ the number of data points, and $P$ the number of periods, which should be an integer to avoid spectral leakage.

We now transform the equidistantly sampled data \eqref{eq:currentVoltageData} into the frequency domain using the discrete Fourier transform (DFT) \cite{cooley1965algorithm},
\begin{align}
    & I(k)=\frac{1}{N} \sum_{n=0}^{N-1} i(n) \euler^{-\frac{j2\pi kn}{N}}& &k=0,1,\hdots,N-1\nonumber\\
    & V(k)=\frac{1}{N} \sum_{n=0}^{N-1} v(n) \euler^{-\frac{j2\pi kn}{N}}& &k=0,1,\hdots,N-1,
    \label{eq:currentVoltageDataFreq}
\end{align}
with index $k$ corresponding to angular frequency $\omega_k=2\pi k/T$, $T=PT_\mathrm{p}$ the total measurement time, and $j$ the imaginary unit ($j^2=-1$). The frequency-domain current and voltage data can be used to calculate impedance and assess linearity and stationarity, as described below.

\section{Experiments}
We now discuss measurements of impedance in a commercial Li-ion battery, comparing single-sine and multisine excitation and exploring several operating conditions. The device under test is a Samsung SDI INR18650-15M cell (Ni-based/graphite), operating between \SI{2.5}{V} and \SI{4.2}{V} with nominal data sheet capacity of \SI{1.5}{Ah}. Experiments were performed in a thermal chamber at \SI{25}{^\circ C} unless otherwise specified.

To determine the actual capacity we first charged the battery at $C/5$ (\SI{300}{mA}) to \SI{4.2}{V} and held the voltage constant until $i(t)<C/40$. We define this operating point as 100\% SOC. The battery was then discharged at $C/5$ until the lower limit of \SI{2.5}{V} was reached. The total current throughput in the discharge cycle is defined as the measured capacity, $Q_\mathrm{meas}=1.522$~Ah.

We then performed impedance measurements under different operating conditions in the frequency range [\SI{20}{mHz}, \SI{1}{kHz}] with 10 logarithmically spaced frequencies per decade, resulting in 48 excited frequencies for the single-sine experiments compared to 40 for the multisine (due to only exciting odd harmonics and loss of resolution at the low frequencies caused by the integer harmonics).

\subsection{Impedance at linear and stationary conditions}

We first measured impedance under steady-state conditions. Starting from 100\% SOC, we discharged the battery in steps of 10\% SOC at $C/5$ (each step duration was \SI{30}{min}) followed by \SI{4}{h} relaxation, single-sine EIS measurement, \SI{10}{min} relaxation, then multisine EIS measurement.
\begin{figure}[t]
    \centering
    \includegraphics[width=\columnwidth]{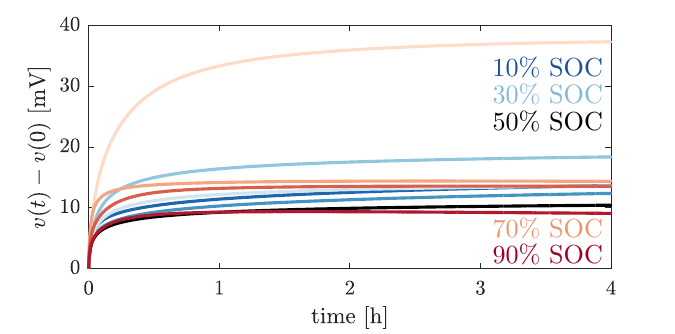}
    \caption{Relaxation voltage for \SI{4}{h} before steady-state EIS measurements. Note that the relaxation time constant depends on SOC.}
    \label{Fig:relaxationVoltage}
\end{figure}
The low discharge currents and long relaxation times allow the battery to reach OCV, and hence stationary conditions, before starting EIS experiments. Note that from the relaxation voltage time series in Fig.~\ref{Fig:relaxationVoltage} the time to reach OCV depends on SOC and a relaxation of over \SI{4}{h} may be required at certain states of charge \cite{fernando2024benchmark}. Moreover, reaching true equilibrium for phase-changing materials like LFP and graphite may require an excessively long relaxation time.

To obtain linear impedance measurements, we used an amplitude of \SI{100}{mA} for the single-sine excitation and a corresponding \SI{70.5}{mA}~rms for the multisine. Transient effects were suppressed by applying four excitation periods and discarding the first period before applying the DFT, resulting in $P=3$ periods (both for single-sine and multisine).

\begin{figure*}
    \centering
    \includegraphics[width=\textwidth]{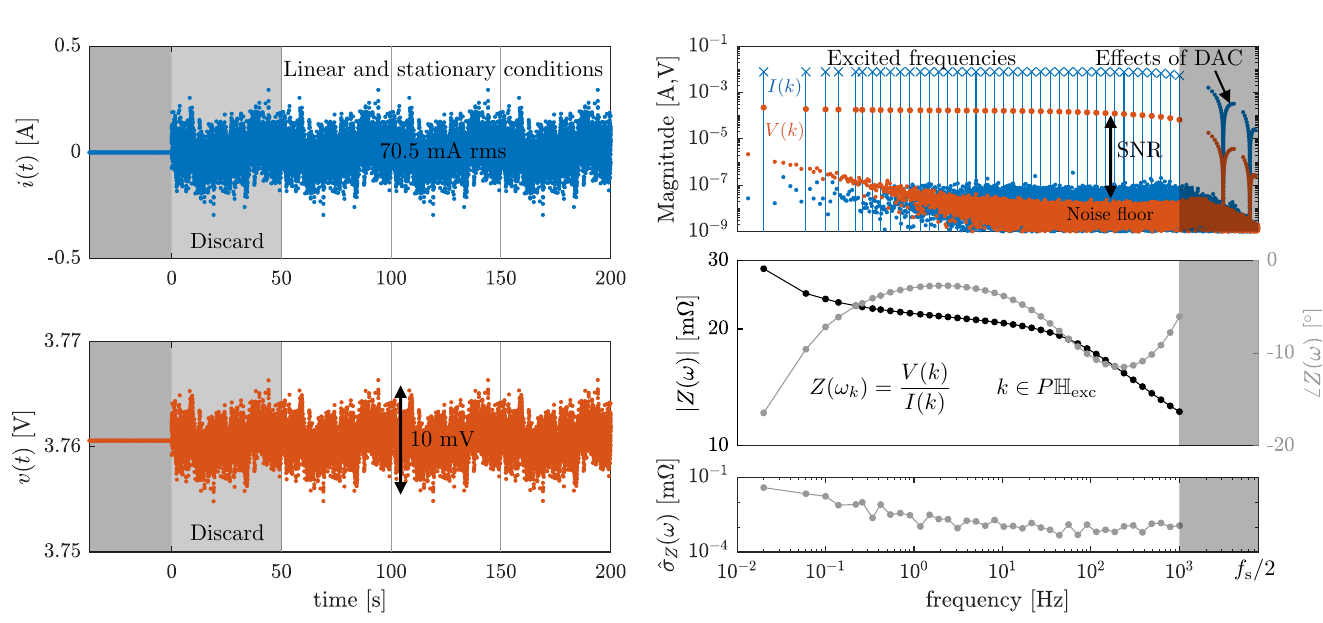}
    \caption{Signals, spectra, and impedance at linear and stationary conditions at 50\% SOC. Time-domain signals were subsampled 125 times for display purposes.}
    \label{Fig:spectraLTI}
\end{figure*}
Fig.~\ref{Fig:spectraLTI} shows the measured multisine data at 50\% SOC. The conditions of linearity and stationarity are indeed satisfied because the non-excited frequencies in the spectra are at the noise floor \cite{hallemans2023electrochemical}. We note an increase of the voltage spectrum at the lower non-excited frequencies which may be due to a small drift signal because OCV was not perfectly reached before the EIS experiment. The SNR is about four orders of magnitude for the voltage and five orders of magnitude for the current, which is more than sufficient for accurate impedance calculation. The steady-state impedance (Fig.~\ref{Fig:spectraLTI}) can be obtained from the current and voltage spectra using
\begin{align}
    Z(\omega_k)=\frac{V(k)}{I(k)} \qquad k\in P\mathbb{H}_\mathrm{exc},
    \label{eq:ZVILTI}
\end{align}
which is a complex quantity depicted on a Nyquist chart by its real part $Z_\mathrm{r}(\omega)$ and imaginary part $Z_\mathrm{j}(\omega)$,  or on a Bode plot with magnitude $\vert Z(\omega)\vert$ and angle $\angle Z(\omega)$ over logarithmic frequency.

In addition to verifying linearity and stationarity, the current and voltage data over several periods enables estimation of a variance $\hat \sigma^2_Z(\omega)$ for the impedance (see Appendix~B and Fig.~\ref{Fig:spectraLTI}), providing \textit{uncertainty bounds}. This is useful for weighting the impedance data for model parametrisation and deciding which parts of the data to trust. Note that this approach works for both multisine and single-sine signals.

\begin{figure}
    \centering
    \includegraphics[width=\columnwidth]{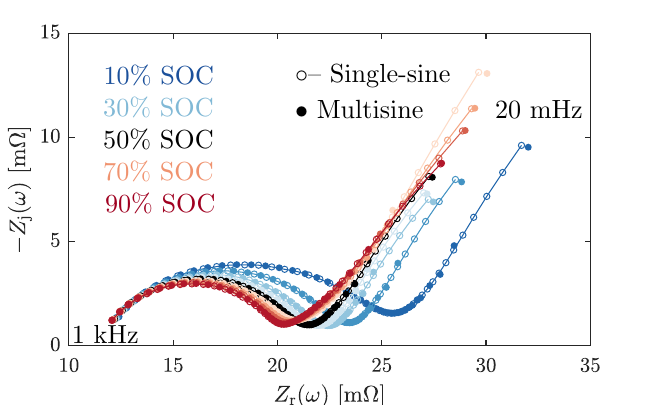}
    \caption{Comparison of single-sine and multisine impedance at linear and stationary conditions and several SOCs. These overlap, but single-sine EIS has better resolution at low frequencies. The excitation magnitude for both single-sine and multisine was \SI{70.5}{mA}~rms.}
    \label{Fig:comparisonSingleSineMultisine}
\end{figure}

Fig.~\ref{Fig:comparisonSingleSineMultisine} shows the measured impedance data at different SOC operating points for both single-sine and multisine excitation. The impedances overlap, confirming that we measure the same quantity. A more detailed discussion of the choice between single-sine and multisine is given in Section~\ref{Section:SingleSineMultisine}. We see how the widths of the charge-transfer arcs and the slopes of the diffusion tails change with SOC.

\subsection{Detecting nonlinearity and nonstationarity}
\label{sec:detect_non}
Artefacts in battery impedance data may lead to incorrect model parametrisation, so it is important to check the validity of measurements. The conditions of linearity and stationarity are mainly determined by the rms value of the excitation, the relaxation time, and temperature stability. Adjusting the rms value of the excitation enables trade-off between SNR and linearity; higher rms gives better SNR but may introduce nonlinearities. As discussed above, stationarity requires the battery to be in equilibrium, necessitating sufficient relaxation time. Stable temperature is also required as battery impedance is very sensitive to temperature \cite{richardson2014battery}.

Most commercial potentiostats only provide estimates of the impedance to users, rather than the underlying measured current and voltage data from EIS experiments, and this limits opportunities to detect nonlinearity and nonstationarity. We now discuss both this situation, and the scenario where direct time-series datasets are available and more accurate methods can be used.

\paragraph{Only impedance available}
When impedance alone is provided by the potentiostat (usually as a table of frequencies vs.\ real and imaginary parts, or magnitudes and phases), nonlinearity and nonstationarity can only be detected using the Kramers-Kronig relations \cite{urquidi1990applications}. However, this method requires integration of the impedance with respect to frequency from zero to infinity, which may not be feasible. Instead, it is commonplace to use a \textit{measurement model} \cite{agarwal1992measurement} that does not aim to understand physical processes, but verifies the conditions of linearity and stationarity by simply trying to fit a plausible model to the data with minimal rms error. A commonly used model is the \textit{Voigt circuit} (a series of parallel-connected resistors and capacitors), which satisfies the Kramers-Kronig relations. Impedance data that can be fitted by this model is then reasonably assumed to satisfy linearity and stationarity requirements.
\begin{figure}[t]
    \centering
    \includegraphics[width=\columnwidth]{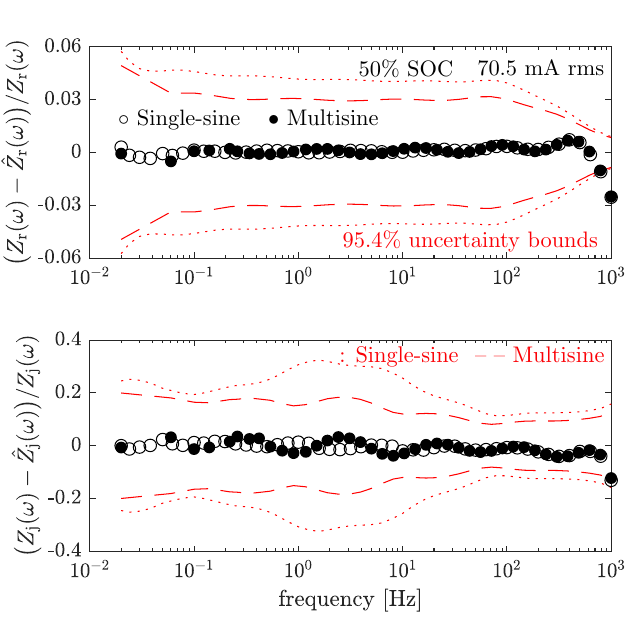}
    \caption{Measurement model residuals (real and imaginary) and 95.4\% uncertainty bounds for both single-sine and multisine at 50\% SOC with \SI{70.5}{mA}~rms excitation. Seven Voigt elements were used for the single-sine data and six for the multisine data.}
    \label{Fig:measurementModel}
\end{figure}

We now verify the validity of the measured single-sine and multisine impedance data of Fig.~\ref{Fig:comparisonSingleSineMultisine} using Orazem's publicly available measurement model software \cite{orazem2024measurement}. The number of statistically significant Voigt elements was chosen by checking that a two-standard deviation (95.4\%) confidence interval for any parameter did not span zero. %
The normalised residuals and 95.4\% uncertainty intervals for the 50\% SOC measurements are shown in Fig.~\ref{Fig:measurementModel}. The multisine data used six Voigt elements while the single-sine data used seven. As the residuals lie within the bounds, the impedance dataset is assumed to satisfy Kramers-Kronig, both for single-sine and multisine. Only the impedance data point at \SI{1}{kHz} lays outside the confidence intervals, showing that either inductive effects are occurring or this data does not satisfy Kramers-Kronig. This point should hence be discarded.

Although impedance data that satisfy linearity and stationarity criteria will satisfy Kramers-Kronig (as in Fig.~\ref{Fig:measurementModel}), the opposite is not necessarily true \cite{urquidi1990applications}. The Kramers-Kronig relations have been shown to be insensitive to failures in linearity \cite{hirschorn2009sensitivity,goh2024comparison}. As pointed out by You et al.\ \cite{you2020application}, multisine data obtained at nonlinear conditions still satisfies Kramers-Kronig through measurement model analysis, because the best linear approximation is fitted \cite{hallemans2023electrochemical}. As we will show later, multisine impedance data obtained at nonstationary conditions may also satisfy Kramers-Kronig. As a result, nonlinearity and nonstationarity of both single-sine and multisine data cannot be verified from impedance data alone, motivating instead the analysis of the raw current and voltage time-series data.

\begin{figure}[t]
    \centering
    \includegraphics[width=\columnwidth]{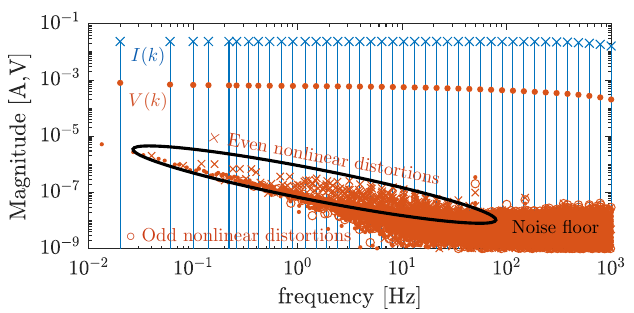}
    \caption{Multisine current and voltage spectra at 10\% SOC for a current excitation with \SI{211.5}{mA}~rms. A distinction is made between even (that are dominant here) and odd nonlinear distortions \cite{hallemans2023electrochemical}.}
    \label{Fig:NLspectra}
\end{figure}

\paragraph{Raw current and voltage data available}
When measured current and voltage data are available, nonlinearity and nonstationarity can easily be detected from the spectra \cite{hallemans2021best,hallemans2023electrochemical}. Indeed, data perfectly satisfying linearity and stationarity and measured for an integer number of periods will show a response at all non-excited frequencies at the noise level (see Fig.~\ref{Fig:spectraLTI}). Nonlinear data will generate harmonics at integer multiples of the fundamental frequencies. This is shown in Fig.~\ref{Fig:NLspectra} where we have applied a \SI{211.5}{mA}~rms excitation, and nonlinearities (crosses) become visible in the spectrum. Nonstationary data will show a drift signal with associated ``skirts'' around the excited frequencies \cite{hallemans2023electrochemical}, as discussed in Section~\ref{Section:operandoEIS}. Nonlinear and nonstationary data will show ``skirts'' not only around the excited frequencies, but also around the nonlinear harmonics \cite{hallemans2021best}.

\begin{figure*}
    \centering
    \includegraphics[width=\textwidth]{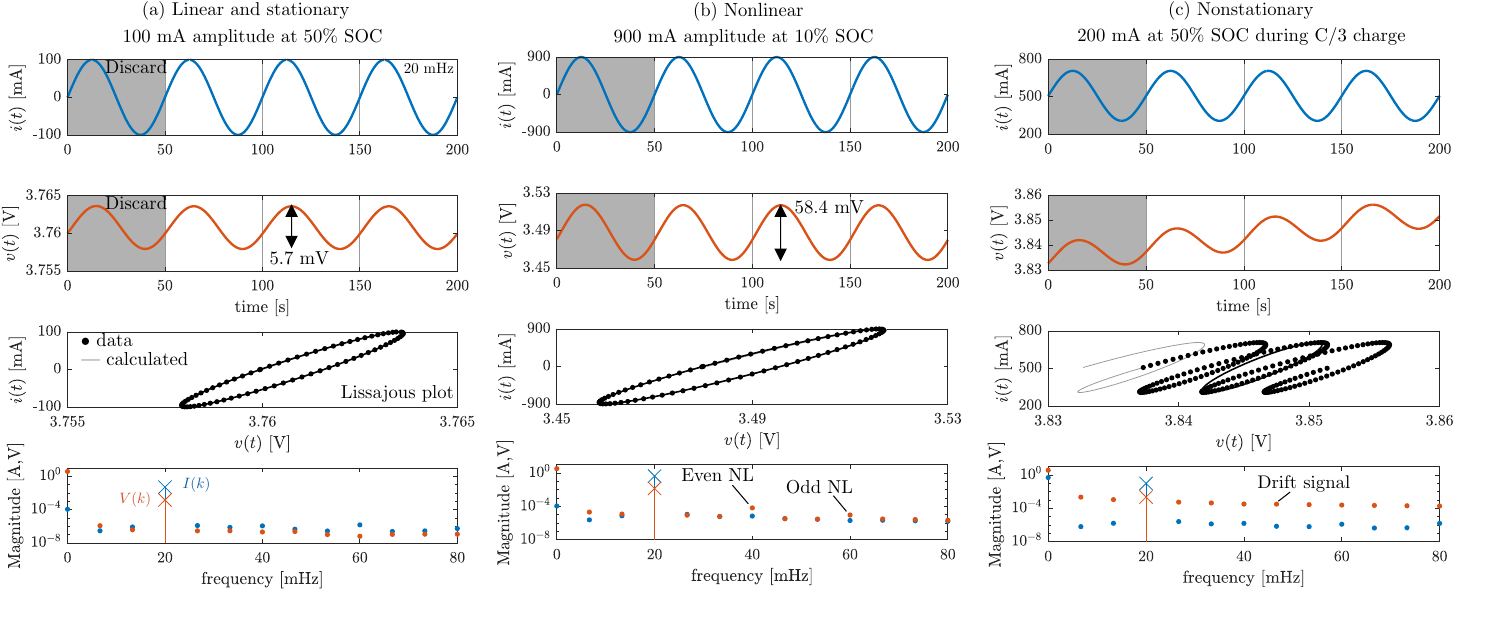}
    \caption{Single-sines at \SI{20}{mHz} at 50\% SOC in linear and stationary conditions measured for four periods of which we discarded the first one. The first period is plotted in grey for the Lissajous plot.}
    \label{Fig:spectraSingleSine}
\end{figure*}
For single-sine excitation, the spectra allow nonlinearity and nonstationarity to be detected in a similar way (see Fig.~\ref{Fig:spectraSingleSine}). The total-harmonic distortion can also be computed for every excited frequency, or Lissajous plots can be used \cite{zabara2024utility}. Both of these approaches were shown by Goh et al.\ \cite{goh2024comparison} to be more sensitive to nonlinear distortions than the measurement model approach discussed above.

\begin{figure*}[htb]
    \centering
    \includegraphics[width=\textwidth]{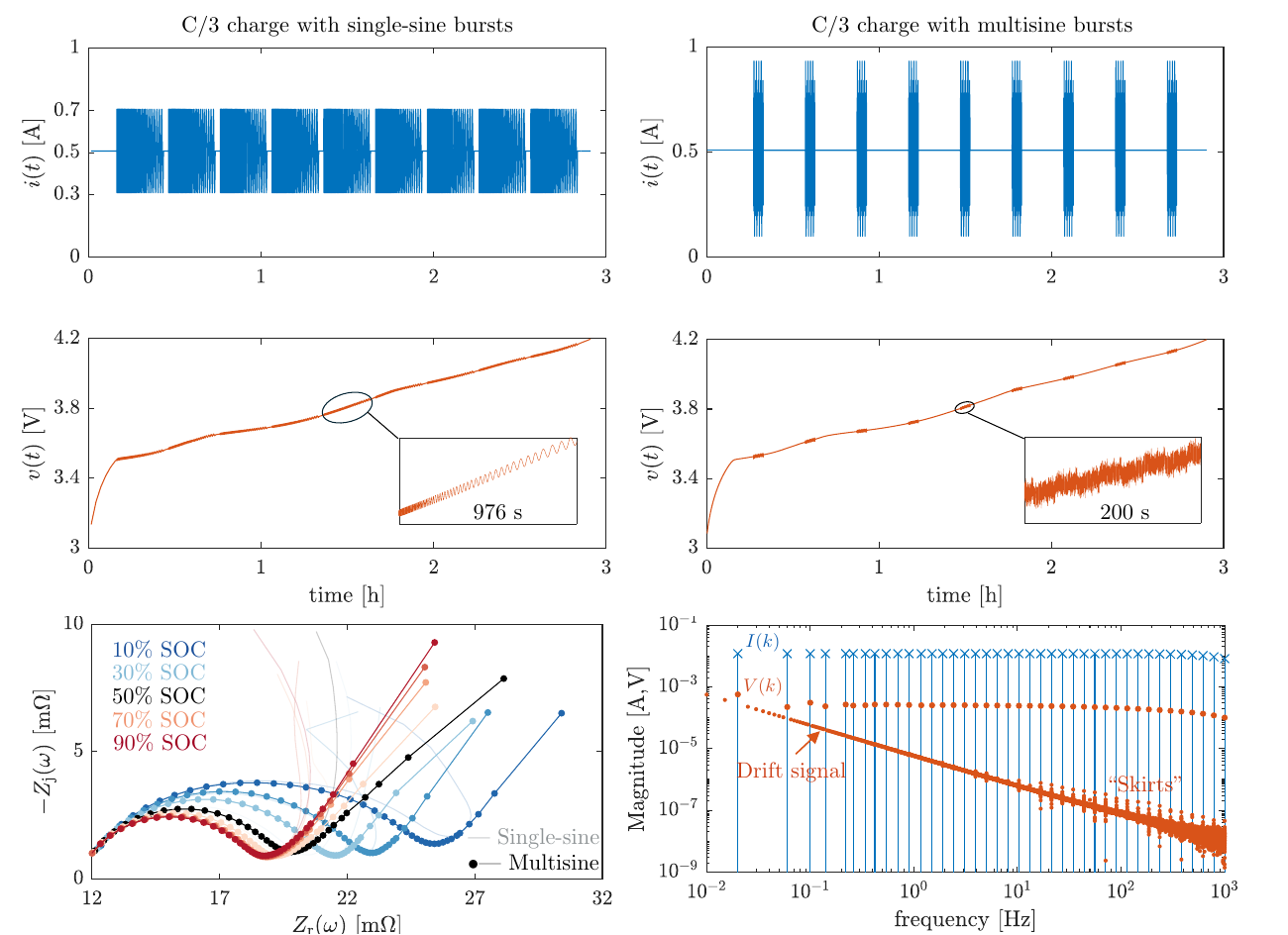}
    \caption{Operando impedance measurements during C/3 charge with single-sine (left) and multisine (right) bursts. The spectra of the multisine current and voltage bursts at 30\% SOC are plotted too, with a drift signal and ``skirts'' present, revealing time-variation during the burst. Single-sine impedance data are computed by dividing the spectra of the sinusoids (without drift removal), and show distortions at low frequencies due to the frequencies being applied at different operating conditions and the voltage drift. Multisine impedance was computed with drift and transient removal \cite{hallemans2022trend}, while not modelling the time-variation.}
    \label{Fig:singleSineOperando}
\end{figure*}

\subsection{Operando EIS}
\label{Section:operandoEIS}
Measuring steady-state impedance, as discussed above, is difficult in practice for batteries due to their long relaxation times and need for temperature stability. Reaching these conditions required approximately \SI{40}{h} of rest in total. This is not merely inconvenient, it indicates that the rest periods of one to two hours commonly used before battery EIS measurements may be insufficient at some states of charge, and that impedance reported as steady-state may in practice retain a dependence on the preceding protocol. %
Instead, it is of interest to measure impedance during normal operations (e.g.\ during charging or relaxation, or while temperature is changing). In these operando conditions, drift signals occur due the OCV and battery impedance changing with SOC and temperature.

Applying single-sines during operation can lead to erroneous impedance data because each sequentially injected frequency is applied at a different operating condition. Instead, because a multisine applies many frequencies simultaneously, it allows correct measurement of operando impedance  \cite{hallemans2023electrochemical,koster2017dynamic,zhu2022operando,hallemans2022operando}. However, obtaining operando impedance from current and voltage spectra requires more involved processing than simple division of voltage and current spectra \eqref{eq:ZVILTI} as these are affected by drift, nonstationarity, and transients that cannot be discarded by removing the first period. Several techniques for resolving time-varying impedance from multisine data are reviewed in Hallemans et al.\ \cite{hallemans2023electrochemical}. Here we apply short excitation bursts (4 periods), and compute the average impedance over the applied multisine window whilst modelling a linear time-variation and removing the drifts and transients using the method of Hallemans et al.\  \cite{hallemans2022operando,hallemans2022trend} that fits basis functions to explain the slow time-evolution of the spectra.

\subsubsection*{Charging and discharging}
To illustrate the challenges associated with \textit{single-sine operando} EIS experiments, we applied single-sine current bursts for four periods (total \SI{976}{s}) superimposed on a $C/3$ charge current centred at 10\% to 90\% SOC, as shown in Fig.~\ref{Fig:singleSineOperando} (top left). Nonstationarity can be detected from the Lissajous plots and spectra, Fig.~\ref{Fig:spectraSingleSine}(c). If the impedance is directly computed by dividing the single-sine spectrum at each excited frequency (without drift removal), then low frequency impedance is affected by the drift signal and errors are introduced because the sinusoids are each applied at different operating conditions, as seen in the bottom left plot of Fig.~\ref{Fig:singleSineOperando}. Measurement models (Section \ref{sec:detect_non}) fail to fit this impedance dataset because it does not satisfy stationarity conditions, and low frequencies must be discarded until a measurement model can be fitted.

One approach for improvement is to remove the drift signal for every voltage sinusoid, for instance by subtracting a straight line between start and end data points (commercial potentiostats often have a drift removal option). However, this should be done carefully as the straight line does not only depend on the drift, but also on the impedance changing over time. In our measurements, even after drift removal, single-sine operando impedance data still did not satisfy Kramers-Kronig through measurement model analysis and low frequencies still had to be discarded. Orazem and Ulgut~\cite{orazem2023use} conclude that drift compensation methods for single-sine experiments may lead to cosmetic improvements but do not usually correct Kramers–Kronig non-compliant data.

On the other hand, a \textit{multisine} excitation allows valid operando impedance data to be measured during this $C/3$ charge over a wider frequency range. The right side of Fig.~\ref{Fig:singleSineOperando} shows multisine bursts around the same operating points as before, and the spectra of one of the bursts in the bottom plot, where we see a drift signal as well as ``skirts'' revealing nonstationarity during the burst \cite{hallemans2023electrochemical}. The resulting impedance data (after drift and transient removal), shown in the bottom left of the same figure, satisfies Kramers-Kronig through the measurement model approach.

\begin{figure}[t]
    \centering
\includegraphics[width=\columnwidth]{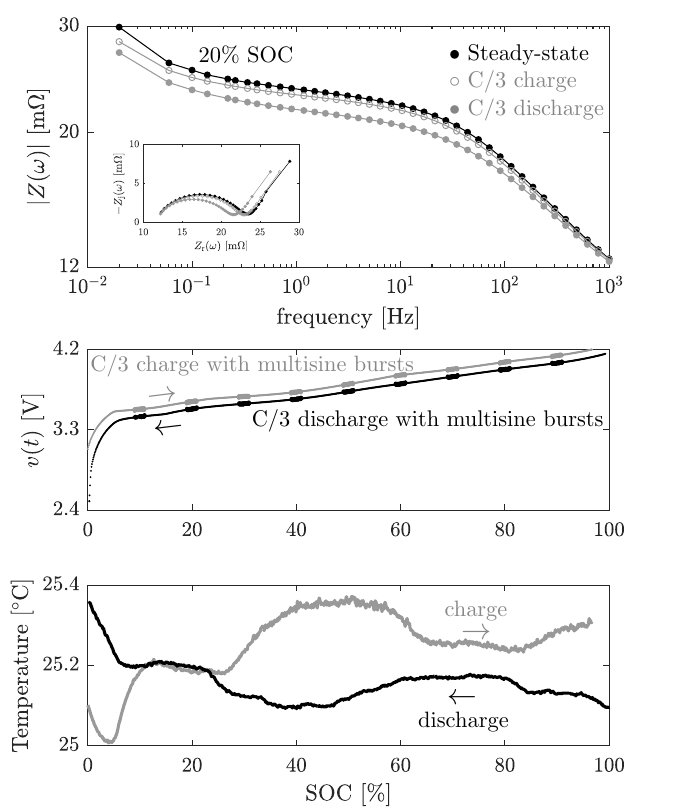}
    \caption{Multisine EIS spectra at steady-state, charge, and discharge, at 20\% SOC.}
    \label{Fig:comparisonImpedanceSteadyStateChargeDischarge}
\end{figure}

We now compare (Fig.~\ref{Fig:comparisonImpedanceSteadyStateChargeDischarge}) impedance data obtained from multisine bursts during $C/3$ charge, $C/3$ discharge, and steady-state, shown here at 20\% SOC. The impedance reported here is the best linear approximation averaged over the excitation burst, during which the SOC changes by 1.9\% (\SI{200}{s} at $C/3$). Because impedance varies smoothly over such a narrow window, this average approximates the impedance at the midpoint SOC, with a residual bias below \SI{0.11}{\milli \ohm} (under 0.6\% of $|Z|$) across the measured band, an order of magnitude smaller than the 0.4--\SI{1.4}{\milli \ohm} difference observed between charge and discharge at matched SOC. %

The impedance decreases slightly during charging and discharging compared to steady state, particularly within the charge-transfer kinetics frequency range (corresponding to the semi-circle in the Nyquist chart). One possible reason could be increased electrolyte conductivity and exchange current densities due to slightly elevated temperatures caused by internal heating during operando conditions \cite{suresh2002temperature}.
However, the temperature increase during the operando experiments is less than \SI{0.4}{\degree C} compared to the steady-state experiment at \SI{25}{\degree C}. Instead, the main reason for the decrease in impedance during operation is that the reaction kinetics (e.g., described by the Butler-Volmer equation) are linearised around a non-zero current, resulting in a different charge-transfer resistance compared to the steady-state case \cite{hallemans2025physics}. We also observe a difference in operando impedance between charge and discharge cases even though the temperature is the same  (\SI{25.2}{\degree C}). %
The reason is the asymmetry of the Butler–Volmer kinetics, which vanishes by construction at equilibrium. Charge and discharge processes also establish concentration gradients with opposite signs within the particles, so the surface concentration, and hence the exchange current density, may differ between the two cases even at equal average SOC. Both mechanisms are inaccessible to a measurement made at rest.

These results have practical consequences for model parametrisation. At equilibrium the exchange current density and the charge-transfer coefficient are combined into a single measured resistance and cannot be separately identified. However, impedance measured at equal and opposite operating currents provides two independent constraints, allowing them to be separated. This offers a route to identifying kinetic parameters that is not available with conventional EIS \cite{huang2015dynamic,hallemans2022operando}, and also enables a direct experimental test for the assumption of symmetric transfer coefficients ($\alpha=$ 0.5) that is common in physics-based battery modelling. Finally, because irreversible heat generation depends on the charge-transfer resistance at the operating current rather than at equilibrium, operando impedance is  directly relevant for quantifying heat generation and creating accurate electro-thermal models.

\begin{figure}[t]
    \centering
\includegraphics[width=\columnwidth]{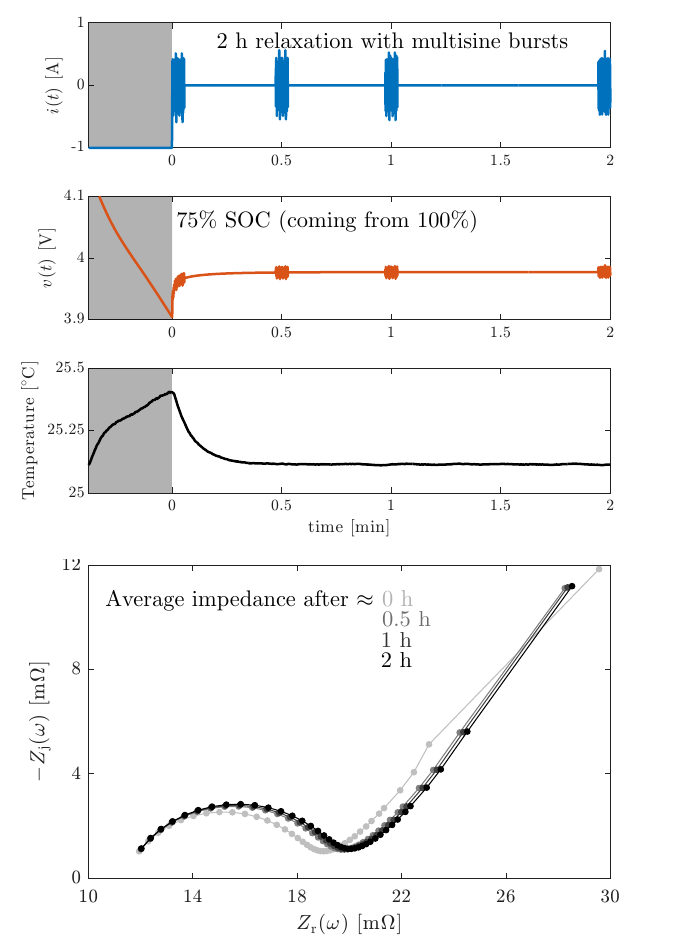}
    \caption{Operando impedance measured in four bursts during relaxation shows a small increase in charge-transfer resistance with time.}
    \label{Fig:relaxation}
\end{figure}

\subsubsection*{Relaxation}
As Fig.~\ref{Fig:relaxationVoltage} shows, battery voltage relaxation to steady-state is a slow process. Measuring the impedance during relaxation might be of interest, for example, to provide characterisation data when an electric vehicle stops.

To explore this, we discharged the battery from 100\% SOC at \SI{-1}{A} to 75\% SOC, then waited for \SI{2}{h} whilst also measuring zero-mean multisine bursts, one at the start of the relaxation, one around \SI{30}{min}, one at \SI{1}{h}, and finally one at the end of the \SI{2}{h} period (Fig.~\ref{Fig:relaxation}). The bursts each consisted of four multisine periods (\SI{200}{s}). We note that the semi-circle becomes larger during relaxation which could be due to the electrolyte and particle surface concentration changing, and also the cell temperature decreasing slightly.

Because each excitation burst yields a full spectrum, the sequence of measurements resolves the rate at which the impedance approaches its equilibrium value; this now becomes a measureable quantity that is a property of the transport processes driving the relaxation rather than of the equilibrium state itself. %
Conventional EIS cannot observe this process, since it requires the relaxation to have finished before measurement can begin. This is particularly relevant for materials in which true equilibrium is slow or ill-defined, such as phase-changing electrodes, where the state probed by a nominally steady-state measurement may depend on how long the cell was rested for.

\subsubsection*{Temperature change}
\begin{figure}[t]
    \centering
\includegraphics[width=\columnwidth]{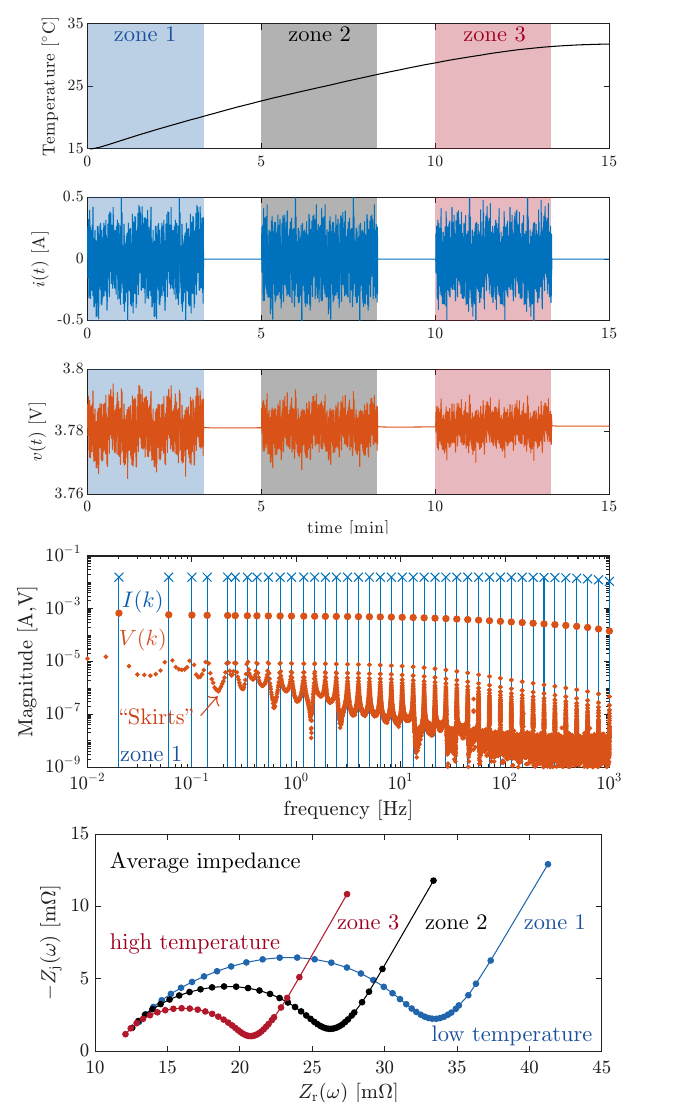}
    \caption{Operando impedance during heating at 50\% SOC. Three multisine bursts were measured, from zone 1 (cold, blue) to zone 3 (warm, red). Voltage spectra in zone 1 show time-variation during the burst as skirts. Average impedance in each zone shows that charge-transfer resistance decreases with temperature, as expected.}
    \label{Fig:temperature}
\end{figure}

Impedance is strongly dependent on temperature in batteries and this can be exploited as an accurate volume-average non-invasive thermal sensor \cite{richardson2015sensorless}. To study the effect of temperature changes on impedance, we started an experiment at 50\% SOC and \SI{15}{\celsius} and at time zero set the reference temperature of the thermal chamber to \SI{30}{\celsius}, leading to a temperature increase. Impedance was measured three times over the \SI{15}{min} experiment, each with a \SI{200}{s} multisine burst followed by \SI{100}{s} relaxation (Fig.~\ref{Fig:temperature}).

The voltage spectra of the first burst (zone~1) measurements show skirts, which are an indication of time-variation during the measurement \cite{hallemans2023electrochemical}, but there is no drift because the OCV does not considerably change over the experiment.
The strong temperature dependence of the mid-frequency impedance, where lower temperatures show larger resistances, is predominantly associated with thermally activated interfacial charge-transfer processes, and reflects the approximately Arrhenius dependence on temperature of the exchange current density \cite{inui2021temperature}. Electrolyte conductivity and diffusivity are also thermally dependent \cite{Hou2020135085}, and could contribute at other frequencies, respectively as a shift in the high-frequency intercept and in the electrolyte diffusion feature, although this is not apparent in the data here.

\section{Single-sine vs.\ multisine impedance}
\label{Section:SingleSineMultisine}

As discussed, both single-sine and multisine excitations have advantages and limitations, and the preferred choice depends on the use case.

\textit{Multisine} experiments are shorter (e.g., \SI{200}{s} vs.\ \SI{976}{s}) and enable measurement during operation. However, very wideband excitations, for instance \SI{100}{\micro Hz} to \SI{1}{MHz}, are infeasible due to memory constraints. Frequency resolution is also lost at lower frequencies due to the constraint of only exciting integer harmonics. Note that, for the cost of not detecting even nonlinearities, the multisine excitation we have presented could produce a better resolution by also exciting even harmonics. We conclude that multisines are especially useful for rapid narrower frequency band measurements that may not be at steady-state.

\textit{Single-sine} experiments on the other hand can be logarithmically spaced, and hence have better resolution at low frequencies. This can be important for model parametrisation \cite{hallemans2025physics}, since features in the impedance might be lost due to an overly sparse set of excited frequencies. Also, SNR might be better for single-sine EIS because more power can be injected at each excited frequency. However, if the battery is not in steady-state, one may have to discard the low frequencies that produce data that do not satisfy conditions of linearity and stationarity. We conclude that single-sines are useful for wideband impedance measurement with a high frequency resolution, in steady-state.
\begin{figure}[t]
    \centering
    \includegraphics[width=\columnwidth]{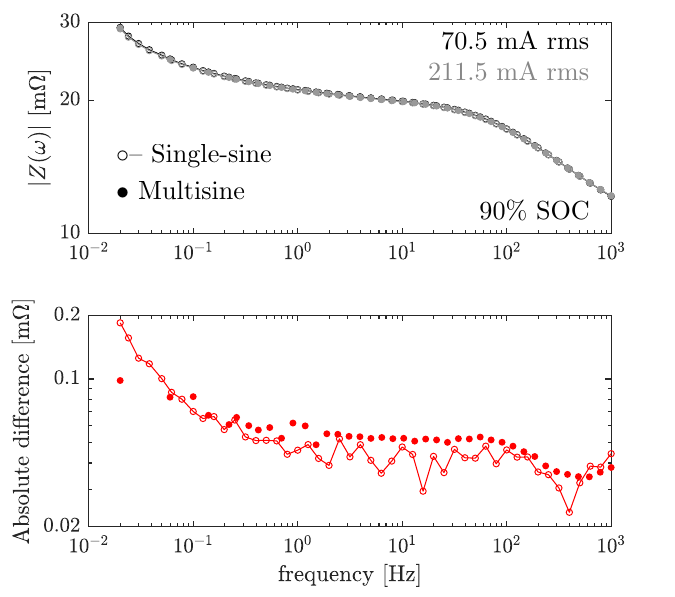}
    \caption{Comparison of single-sine and multisine impedance data at 90\% SOC for different excitation amplitudes. Black and grey nearly overlap on the top figure.}
    \label{Fig:comparisonAmplitude}
\end{figure}

For EIS measurements at equivalent rms current, more nonlinearity will typically be present for a multisine as the excitation amplitude span is larger due to the higher crest factor \eqref{eq:crestfactor}. However, we obtained very similar impedance results for single-sine and multisine even when increasing rms current from \SI{70.5}{mA}~rms to \SI{211.5}{mA}~rms (Fig.~\ref{Fig:comparisonAmplitude}). Although even nonlinearities are introduced in the \SI{211.5}{mA}~rms experiment (Fig.~\ref{Fig:NLspectra}), these do not affect the measured impedance because only odd frequencies were excited \cite{hallemans2023electrochemical}, and, hence we obtain nearly the same impedance as for the smaller \SI{70.5}{mA}~rms experiment.
\begin{figure}[t]
    \centering
    \includegraphics[width=\columnwidth]{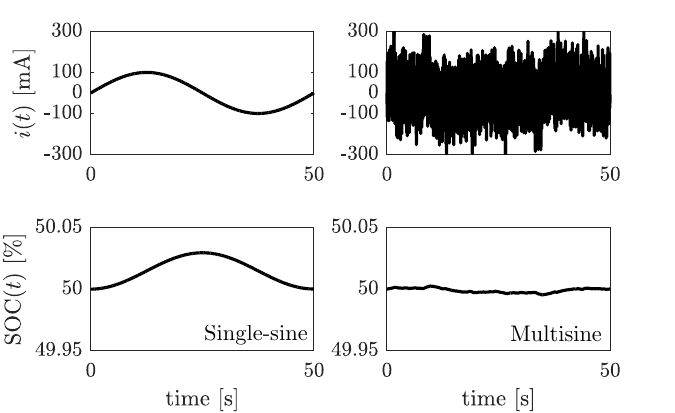}
    \caption{Variation of SOC for single-sine vs.\ multisine at equivalent rms current.}
    \label{Fig:SOCVariation}
\end{figure}

Another observation is that the SOC fluctuations,
\begin{align}
    \text{SOC}(t)=\text{SOC}_0+ \frac{100}{3600 Q_\text{meas}}\int_0^t i(\tau)\mathrm{d}\tau \qquad
    \nonumber \\ t\in [0,T_\mathrm{p}],
\end{align}
are significantly smaller during the multisine experiment compared to the lowest frequency  single-sine test (Fig.~\ref{Fig:SOCVariation}). At equal rms current, the SOC excursion during the multisine excitation is 0.0074\% peak-to-peak, compared with 0.0294\% for the \SI{20}{mHz} single-sine, a factor of 3.9 smaller.
This is because a multisine alternates many times between charging and discharging, while a single-sine charges for half the period and discharges for the other half. For single-sines, this SOC swing typically limits the lowest measurable frequency to avoid nonlinear distortion. Also, for single-sine excitation the SOC swing is different at every frequency---with the lowest frequency showing the largest swing---while for a multisine the SOC swing is the same for all frequencies. Hence, for single-sine excitation the linearisation occurs over a different SOC width for every frequency, which may explain the small discrepancy between the single-sine and multisine impedance data at low frequencies due to the nonlinear OCV-SOC function. Note for both multisine and single-sine, the amplitudes can be chosen to be equal over all frequencies, or alternatively shaped based on the noise floor and impedance characteristics.

\section{Conclusions}
In this paper we have discussed the design and measurement of multisine electrochemical impedance data with a commercial potentiostat. We have demonstrated measured impedance on Li-ion cells in steady-state, showing that single-sine and multisine experiments with the same rms excitation current measure the same impedance. We have discussed how to detect nonlinearity and nonstationarity, and why it is useful to analyse the current and voltage time-series data directly, especially in the frequency domain.

Several examples were explored demonstrating the benefits of multisine for measuring operando impedance in Li-ion batteries. We measured multisine impedance during charge, discharge, relaxation, and temperature changes. Impedance during battery charging and especially discharging is different to steady-state impedance, reflecting asymmetry in the kinetics that cannot be seen with conventional methods. Understanding this behaviour is useful, for example, for quantifying the reaction heat generation. Impedance measurements during relaxation give additional insight into transport, and measurements during temperature changes allow characterisation of  thermal dependence of exchange current and other properties.

Overall, the value of multisine operando EIS lies both in the time it saves and in what it makes measurable by providing access to non-equilibrium behaviour. Impedance measured under load, during relaxation, or during a temperature transient, characterises a device in states that conventional EIS must exclude. Measurements during relaxation are a case in point: because the impedance is resolved repeatedly as the cell approaches equilibrium, the rate of that approach becomes observable. This is not possible with a technique that requires relaxation to be complete before measurement begins.

Although we have focused on batteries, the argument applies wherever the operating state differs from the rest state. Fuel cells and electrolysers are conventionally characterised by impedance around steady operating points under substantial DC load; multisine operando impedance methods extend this capability to conditions in which the operating point evolves, for example during load changes, start-up, changes in reactant conditions, or progressive degradation. Similar opportunities arise in corrosion systems with evolving interfaces and in electrochemical reactors undergoing changes during continuous operation. However, we emphasise again that potentiostats should make the measured raw current and voltage data from EIS experiments available to users, since these would enable users to assess the conditions of linearity and stationarity and undertake measurements away from steady-state conditions.

\section*{Appendix A: Measurement protocols}
The following pseudo-code sections give the test protocols used in this paper. Further details (e.g., frequency ranges) are given in the main text.
\begin{lstlisting}[caption=Classical EIS protocol, numbers=none, numbers=none,keywordstyle=\color{black}]
Start at 100% SOC in steady-state
  Steps:
    0: Discharge at C/5 for 30 minutes
    1: Rest for 3 hours
    2: Single-sines for 976 seconds
    3: Rest for 10 minutes
    4: Multisine for 200 seconds
    5: Rest for 10 minutes
  Cycle:
    Start: 0
    End: 5
    Count: 9
\end{lstlisting}
\begin{lstlisting}[caption=Charge multisine EIS protocol, numbers=none, numbers=none,keywordstyle=\color{black}]
Start at 0% SOC in steady-state
  Steps:
    0: Charge at C/3 for 980 seconds
    1: Charge at C/3 + multisine for 200 seconds
    2: Charge at C/3 for 880 seconds
    3: Charge at C/3 + multisine for 200 seconds
    4: Charge at C/3 until 4.2 V
  Cycle:
    Start: 2
    End: 3
    Count: 8
\end{lstlisting}
\begin{lstlisting}[caption=Charge single-sine EIS protocol, numbers=none, numbers=none,keywordstyle=\color{black}]
Start at 0% SOC in steady-state
  Steps:
    0: Charge at C/3 for 592 seconds
    1: Charge at C/3 + single-sine for 976 seconds
    2: Charge at C/3 for 104 seconds
    3: Charge at C/3 + single-sines for 976 seconds
    4: Charge at C/3 until 4.2 V
  Cycle:
    Start: 2
    End: 3
    Count: 8
\end{lstlisting}
\begin{lstlisting}[caption=Discharge multisine EIS protocol, numbers=none, numbers=none,keywordstyle=\color{black}]
Start at 100% SOC in steady-state
  Steps:
    0: Discharge at C/3 for 980 seconds
    1: Discharge at C/3 + multisine for 200 seconds
    2: Discharge at C/3 for 880 seconds
    3: Discharge at C/3 + multisine for 200 seconds
    4: Discharge at C/3 until 2.5 V
  Cycle:
    Start: 2
    End: 3
    Count: 8
\end{lstlisting}
\begin{lstlisting}[caption=Relaxation EIS protocol, numbers=none, numbers=none,keywordstyle=\color{black}]
Start at 100% SOC in steady-state
  Steps:
    0: Discharge at -1 A until 75% SOC
    1: Multisine for 200 seconds
    2: Rest for 1500 seconds
    3: Multisine for 200 seconds
    4: Rest for 1600 seconds
    5: Multisine for 200 seconds
    6: Rest for 3300 seconds
    7: Multisine for 200 seconds
\end{lstlisting}
\begin{lstlisting}[caption=Temperature change protocol, numbers=none,keywordstyle=\color{black}]
Start at 50% SOC and set temperature to 30 degrees C (initially 15)
  Steps:
    0: Multisine EIS (200 seconds)
    1: Rest for 100 seconds
  Cycle:
    Start: 0
    End: 1
    Count: 3
\end{lstlisting}

\section*{Appendix B: Impedance uncertainty}
The variance $\hat \sigma^2_Z(\omega)$ of the impedance can be estimated from several measured periods of a multisine (or single-sine) in steady-state.

Denote $x^{[p]}(n)$ as data from the $p$-th period (of total $P$ periods), with $x=\{i,v\}$ and $n=\{0,1,\hdots,N_\mathrm{p}-1 \}$. In the frequency domain,
\begin{align}
    X^{[p]}(k)=\frac{1}{N_\mathrm{p}} \sum_{n=0}^{N_\mathrm{p}-1} x^{[p]}(n) \euler ^{-\frac{j2\pi kn}{N_\mathrm{p}}} \nonumber \\ k=0,1,\hdots,N_\mathrm{p}-1.
\end{align}
Next, average the spectra over the $P$ measured periods,
\begin{align}
    \hat X(k)=\frac{1}{P}\sum_{p=1}^P X^{[p]}(k).
\end{align}
Note that $\hat X(k)=X(Pk)$ if $X(k)$ is the DFT over all $P$ periods, showing that taking the DFT over the total dataset record averages automatically. However, from the DFTs of the individual periods we can also compute variances
\begin{align}
    \sigma_X^2(k)=\frac{1}{P(P-1)}\sum_{p=1}^P\big\vert X^{[p]}(k)-\hat X(k)\big\vert^2 ,
\end{align}
and a cross-variance
\begin{align}
    \sigma_{VI}^2(k)=\frac{1}{P(P-1)}\sum_{p=1}^P\big(V^{[p]}(k)\nonumber \\ - \hat V(k)\big)\overline{\big(I^{[p]}(k)-\hat I(k)\big)},
\end{align}
and finally calculate the variance of the impedance \cite{pintelon2012system}
\begin{align}
    \hat \sigma^2_Z(\omega_k)=\vert Z(\omega_k)\vert^2\Bigg(\frac{\sigma^2_V(k)}{\vert \hat V(k)\vert^2}+\frac{\sigma^2_I(k)}{\vert \hat I(k)\vert^2}- \nonumber \\ 2\mathrm{Re}\Bigg(\frac{\sigma^2_{VI}(k)}{\hat V(k) \overline{\hat I(k)}}\Bigg)\Bigg),
\end{align}
with $\omega_k=2\pi k/T_\mathrm{p}$ and $k\in \mathbb{H}_\mathrm{exc}$. Assuming the noise in the time domain is normally distributed, this leads to circular confidence intervals in the complex plane with radius $\sqrt{-\log(1-a)}\hat \sigma_Z(\omega_k)$ for a $100a$\% confidence region. For a 95\% confidence region, for instance, we obtain radii of approximately $\sqrt{3}\sigma_Z(\omega_k)$.

\section*{Acknowledgements}
This work dedicated to Rik Pintelon was supported by the Faraday Institution Nextrode (FIRG066) and Multiscale Modelling (FIRG059) projects, and EU IntelLiGent (101069765) project via the UKRI Horizon Europe Guarantee (10038031). We thank Mark E.\ Orazem for his useful comments about this work. We acknowledge that our measured multisine impedance data in previous works \cite{hallemans2023electrochemical,hallemans2022operando} contains aliasing errors above \SI{10}{Hz} due to the lack of an anti-alias filter and oversampling. 

\section*{Data availability}
The measured current and voltage datasets underlying this study are openly available at Oxford Research Archive under a CC BY 4.0 licence, DOI: 10.5287/ora-2a8zazmqr \cite{dataset2026}.

\section*{Competing Interests}
David Howey is a co-founder of Brill Power. Peter Keil is the founder of Battery Dynamics, and Heiko Seel-Mayer is a Development Engineer at Battery Dynamics. Noël Hallemans now works at Breathe Battery Technologies. The other authors have no competing interests to declare.

\bibliographystyle{elsarticle-num}
\bibliography{eis}

\end{document}